\input{aipcheck}

\documentclass[
    ,final            
  ]
  {aipproc}

\layoutstyle{8x11single}
\newcommand{\be}{\begin{equation}}
\newcommand{\ee}{\end{equation}}
\newcommand{\bea}{\begin{eqnarray}}
\newcommand{\eea}{\end{eqnarray}}
\newcommand{\bref}[1]{(\ref{#1})}
\newcommand{\la}{\langle}
\newcommand{\ra}{\rangle}

\newcommand{\pa}{\partial}

\begin{document}

\title{Minimal SO(10) GUT in 4D and its extension to 5D \footnote{This is a talk at GUT2012 held on March 15-17 at Kyoto.}}


\author{Takeshi FUKUYAMA}{
  address={Department of Physics and R-GIRO, Ritsumeikan University,\\
 Kusatsu, Shiga,
525-8577, Japan}
}

\begin{abstract}
 The problems of renormalizable minimal SUSY SO(10) GUT in 4D are discussed. 
Its highly predictivity has been charged with many observations, which urges further progresses. We show why and how broad data fittings and conceptual problems drive us to 5D and how it improves the model.
\end{abstract}

\maketitle


\section{Minimal Supersymmetric SO(10) GUT}
SUSY GUT is the most promising candidate beyond the Standard Model (SM).  
The SM is a very powerful theory but it has the application limit like the other great theories.
Among many SUSY GUT models, a renormalizable minimal SUSY SO(10) GUTs (minimal SO(10) GUTs) heve been considered to be very promising because of their high predictivity. Minimal implies that ${\bf 10}$ and $\overline{{\bf 10}}$ Higgs are incorporated into Yukawa coupling. This model was first applied to neutrino oscillation data by \cite{Babu}.
Since that time, we have developed the following critical points among th other groups. 
\begin{itemize}
\item
The phase factors were proved to be indispensable for the neutrino oscillation data \cite{Fuku1}, 
\item
RGE effect was incorporated, which enables us to match up with the low energy data from GUT relations \cite{Fuku2}.
\item
The complete symmetry breaking pattern from GUT to the SM was shown \cite{Fuku3} etc.
\end{itemize}
Yukawa coupling is given by
\begin{eqnarray}
 W_Y = Y_{10}^{ij} {\bf 16}_i H_{10} {\bf 16}_j 
           +Y_{126}^{ij} {\bf 16}_i H_{126} {\bf 16}_j \; , 
\label{Yukawa1}
\end{eqnarray} 
where ${\bf 16}_i$ is the matter multiplet of the $i$-th generation,  
 $H_{10}$ and $H_{126}$ are the Higgs multiplet 
 of {\bf 10} and $\overline{\bf 126} $ representations 
 under SO(10), respectively. 
Providing the Higgs VEVs, 
 $H_u = v \sin \beta$ and $H_d = v \cos \beta$ 
 with $v=174 \mbox{GeV}$, 
 the quark and lepton mass matrices can be read off as%
\begin{eqnarray}
\label{massmatrix}
  M_u &=& c_{10} M_{10} + c_{126} M_{126},~~   
  M_d =     M_{10} +     M_{126}   \nonumber \\
  M_D &=& c_{10} M_{10} -3 c_{126} M_{126},~~
  M_e =    M_{10} -3     M_{126}   \nonumber \\
  M_L &=& c_L M_{126},~~ 
  M_R = c_R M_{126}  \nonumber \; , 
 \end{eqnarray} 
where $M_u$, $M_d$, $M_D$, $M_e$, $M_T$, and $M_R$ 
 denote the up-type quark, down-type quark, 
 Dirac neutrino, charged-lepton, left-handed Majorana, and 
 right-handed Majorana neutrino mass matrices, respectively. 

In \cite{Fuku1} and \cite{Fuku2}, we set $c_L=0$ and $c_R$ is real (type I seesaw). We do not discuss Type II seesaw dominant model simply because of lack of space.

\begin{table}
\caption{The input values of $\tan \beta$, $m_s(M_Z)$
and $\delta$ in the CKM matrix and the outputs for the neutrino
oscillation parameters.}
{\begin{tabular}{c|cc|c|ccc|c}
\hline \hline
 $\tan \beta $ & $m_s(M_Z)$ & $\delta$  & $\sigma $
 & $\sin^2 2 \theta_{1 2}$
 & $\sin^2 2 \theta_{2 3}$
 & $\sin^2 2 \theta_{1 3} $
 & $\Delta m_{\odot}^2/\Delta m_{\oplus}^2$ \\ \hline
40 & 0.0718 & $ 93.6^\circ $ & 3.190& 
0.738 & 0.900 & 0.163 & 0.205 \\
45 & 0.0729 & $ 86.4^\circ $ & 3.198& 
0.723 & 0.895 & 0.164 & 0.188 \\
50 & 0.0747 & $ 77.4^\circ $ & 3.200& 
0.683 & 0.901 & 0.164 & 0.200 \\
55 & 0.0800 & $ 57.6^\circ $ & 3.201& 
0.638 & 0.878 & 0.152 & 0.198 \\
\hline \hline
\end{tabular}}
\end{table}
Together with real $c_R$ which is used to determine the overall neutrino mass scale,  this system fixes all mass matrices, very strong predictability to the fermion mass matrices.
The reasonable results we found are listed in Table 1.
Thus we can fix neutrino mixing angles, abosulute neutrino masses, four CP phases (one in the CKM and three in the MNS matrices). Moreover it fixes Dirac $M_D$ and $M_R$. The former (latter) is crucial for lepton flavour violation (leptogenesis mainly via $M_R$ decay). 
In the basis where both of the charged-lepton 
 and right-handed Majorana neutrino mass matrices 
 are diagonal with real and positive eigenvalues, 
 the neutrino Dirac Yukawa coupling matrix at the GUT scale 
 is found to be \footnote{We are now reconsidering data fitting with
the update experimental data and new RGE results. It gives little bit differen values from \bref{Ynu} but the LFV results are not essentially changed.}
\begin{eqnarray}
 Y_{\nu} = 
\left( 
 \begin{array}{ccc}
-0.000135 - 0.00273 i & 0.00113  + 0.0136 i  & 0.0339   + 0.0580 i  \\ 
 0.00759  + 0.0119 i  & -0.0270   - 0.00419  i  & -0.272    - 0.175   i  \\ 
-0.0280   + 0.00397 i & 0.0635   - 0.0119 i  &  0.491  - 0.526 i 
 \end{array}   \right) \; .  
\label{Ynu}
\end{eqnarray}     
LFV effect most directly emerges 
 in the left-handed slepton mass matrix 
 through the RGEs such as \cite{Hisano-etal}
\begin{eqnarray}
\mu \frac{d}{d \mu} 
  \left( m^2_{\tilde{\ell}} \right)_{ij}
&=&  \mu \frac{d}{d \mu} 
  \left( m^2_{\tilde{\ell}} \right)_{ij} \Big|_{\mbox{MSSM}} 
 \nonumber \\
&+& \frac{1}{16 \pi^2} 
\left( m^2_{\tilde{\ell}} Y_{\nu}^{\dagger} Y_{\nu}
 + Y_{\nu}^{\dagger} Y_{\nu} m^2_{\tilde{\ell}} 
 + 2  Y_{\nu}^{\dagger} m^2_{\tilde{\nu}} Y_{\nu}
 + 2 m_{H_u}^2 Y_{\nu}^{\dagger} Y_{\nu} 
 + 2  A_{\nu}^{\dagger} A_{\nu} \right)_{ij}  \; ,
 \label{RGE} 
\end{eqnarray}
where the first term in the right hand side denotes 
 the normal MSSM term with no LFV. 
We have found $Y_\nu$ explicitly and we can calculate LFV and
 related phenomena unambiguously \cite{Fukuyama2}

It also gives proton decay ratio unambiguously \cite{Fuku4}.

%


It is important that this data fitting was essentially good before Kamland data appeared \cite{Kamland} except for fast proton decay \cite{Fuku7}.
After Kamland, the fitting is not good for $\theta_{13}$ and $\frac{\Delta m_{atm}^2}{\Delta m_{sol}^2}$.
However, this data fitting was performed to show how minimal SO(10) GUT is predictive, and we have not exhausted parameter searching.

On the other hand, it has been long expected to uncover the symmetry breaking pattern from GUT to the SM.
 The simplest Higgs superpotential at the renormalizable level
is given by \cite{clark}, \cite{lee}, \cite{aulakh}
\begin{equation}
W=m_1 \Phi^2 + m_2 \Delta \overline{\Delta} 
+m_3 H^2
+\lambda_1 \Phi^3 + \lambda_2 \Phi \Delta \overline{\Delta}
+\lambda_3 \Phi \Delta H + \lambda_4 \Phi \overline{\Delta} H \;,
\label{lee}
\end{equation}
where $\Phi ={\bf 210}$, $\Delta ={\bf 126}$, 
$\overline{\Delta} ={\bf \overline{126}}$ and $H={\bf 10}$.
The interactions of ${\bf 210}$, ${\bf \overline{126}}$, 
${\bf 126}$ and ${\bf 10}$ lead to some complexities 
in decomposing the GUT representations to the MSSM 
and in getting the low energy mass spectra.  
Particularly, the CG coefficients 
corresponding to the decompositions of 
${\rm SO}(10) \to 
{\rm SU}(3)_C \times {\rm SU}(2)_L \times {\rm U}(1)_Y$ 
have to be found.
This problem was first attacked by X.~G.~He and S.~Meljanac
\cite{he} and further by J.Sato \cite{Joe} and D.~G.~Lee \cite{lee}. 
But they did not present the explicit form of mass matrices for 
a variety of Higgs fields and also did not perform a formulation 
of the proton life time analysis.  This is very labourious work and it is indispensable for the data fit of low energy physics.
We completed that program in \cite{Fuku3} (See also \cite{Bajc:2004xe}, \cite{Aulakh:2004hm}, \cite{FukuHiggs}).
This construction is only possible for the minimal SO(10) GUT. So far many models have suggested the intermediate energy scales between GUT and the SM like seesaw scale and Peccei-Quinn symmetry breaking scale etc. The minimal SO(10) GUT explicitly gives these intermediate energy scales. However, these scales give rise to a trouble in the gauge coupling unification 
 \cite{Bertolini:2006pe}. 
Thus we have mainly two problems; one is on the data fitting and another is on the gauge coupling unification.

\section{Problems of minimal SO(10) GUT}
\subsection{model modifications in 4D}
First we consider on the improvement of data fitting.
More eraborate parameter searching including type II seesaw ($c_L\neq 0$) was done by \cite{Babu-Macesanu}. See also \cite{Babu1} incorporating the recent Daya-Bay result \cite{Daya-Bay}.
Another approach is to add ${\bf 120}$ Higgs \cite{Fuku6} where parameteters are increased and data fitting is improved and fast proton decay also remedied.
Since ${\bf 120}$ has two SM doublets $({\bf 1,2,2})$ and $({\bf 15,2,2})$, mass matrices become 
\bea
M_u&=&c_{10}M_{10}+c_{120}^{(1)}M_{120}+c_{126}M_{126},~~
M_d=M_{10}+M_{120}+M_{126}\nonumber\\
M_D&=&c_{10}M_{10}+c_{120}^{(2)}M_{120}-3c_{126}M_{126},~~
M_e=M_{10}+c_{120}^{(3)}M_{120}-3M_{126}\\
M_L&=&c_LM_{126},~~
M_R=c_RM_{126}\nonumber
\label{mass2}
\eea
Here
\be
c_{120}^{(1)}=\frac{\la \phi_+\ra+\la \phi'_+\ra}{-\la \phi_-\ra+\la \phi'_-\ra},~~c_{120}^{(2)}=\frac{\la \phi_+\ra-3\la \phi'_+\ra}{-\la \phi_-\ra+\la \phi'_-\ra},~~c_{120}^{(3)}=\frac{-\la \phi_-\ra-3\la \phi'_-\ra}{-\la \phi_-\ra+\la \phi'_-\ra},
\ee
where
$\la\phi_{\pm}\ra$ are expectation values of ${\bf (1,2,2)}$ of ${\bf 120}$, and$\la\phi'_{\pm}\ra$ are those of ${\bf (15,2,2)}$ of ${\bf 120}$.

This model has been extensively explored by \cite{D-M-M}. 
In the original model, ${\bf 126}$ takes part of Majorana neutrinos, as well as charged fermions \bref{massmatrix}. In other word, $Y_{126}$ was of $O(1)$ as $Y_{10}$ to recover the wrong SU(5) mass relation $M_e=M_d$.

The mass of heavy right handed Majorana neutrino is surely several orders smaller than $M_{GUT}$ (we recognised that type II seesaw is subdominant), which means that we are forced to have the vev $v_R$ of intermediate energy scale.
However, we have additionally many parameters and can use ${\bf 126}$ for determing $M_R$ and $M_L$ independently on the determination of charged fermion mass matrices. That is $Y_{126}$ is free from order one unlike the minimal case and vevs are free from having the intermediate energy scales and we may remedy the gauge coupling crisis mentioned later. This seems to  be fine at least for data fittings of low energy.
 
The reason why the gauge coupling unification is broken is as follows.
The renormalizable SUSY GUT with Higgs fields of high dimensional representation has many Standard Model vacua. However such intermediate energy scale is fixed by only single parameter as was shown from vev conditions of the Higgs superpotential \bref{lee} also 
\be
\frac{c_{10}}{c_{126}}=-\frac{3(v-1)(v+1)(2v-1)(v^3+5v-1)}{8v^6-27v^5+38v^4-70v^3+87v^2-31v+3},
\ee
$v\equiv \frac{\phi_3}{{\cal M}_1}$ with $\phi_3=(12+34)(56+78+90)$ and ${\cal M}_1=
12\left(\frac{m_1}{\lambda_1}\right)$ (See \cite{Fuku3} for notation).

So if we add another Higgs, if we retain renormalizability, ${\bf 120}$, then by virtue of ${\bf 120}$, $\frac{c_{10}}{c_{126}}$ can be free \cite{Mimurap}.

However, it seems to be very difficult to recover gauge couling renormalizability even in this case since there still remain four intermediate energy scales.

There are the other conceptual problems which become the obstavcle towards the complete GUT in 4D.

The great advantage of minimal SO(10) model was its high predictivity, implying that all quark-leptons mass matrices including Dirac and Majorana neutrinos, are completely determined.

In order that such theory becomes the SM of next generation, we must also study Doublet-Triplet problem and SUSY breaking mechanism. We will see this point soon later.

One of the other approaches is to use Split Susy \cite{Bajcsplit} with light gauginos and higgsinos in 100 TeV range and superheavy squarks and sleptons in energy scale close to GUT. However, it is essentially non SUSY and unnatural.

Of course, there is a choice of adopting nonsusy SO(10) GUT \cite{Malinsky}.
\subsection{NO-GO theorem in 4D}
However, there is arguments that it is impossible to construct a GUT in 4D with a finite number of multiplets that leads to the MSSM with a residual R symmetry \cite{Ratz}, whose NO GO theorem is not applicable to extra dimensions.
Let me explain this:
SUSY invariant action is assumed to be invariant under global $U(1)_R$ transformation (for N=1 supersymmetry as we consider in this review),
\be
\theta\rightarrow e^{i\alpha}\theta,~~\theta^\dagger\rightarrow e^{-i\alpha}\theta^\dagger,
\label{U1R}
\ee
impling that R-charge of $\theta$ and $\theta^\dagger$ are 1 and -1, respectively.
\be
\Phi=\phi(y)+\sqrt{2}\theta\psi(y)+\theta\theta F(y)
\ee
with
\be
y^\mu\equiv x^\mu+i\theta^\dagger\overline{\sigma}^\mu\theta.
\ee
Vector superfild is real and its R-charge =0. Vector superfield in Wess-Zumino gauge is
\be
V=\theta^\dagger\overline{\sigma}^\mu\theta A_\mu+\theta^\dagger\theta^\dagger\theta\lambda+\theta\theta\theta^\dagger\lambda^\dagger+\frac{1}{2}\theta\theta\theta^\dagger\theta^\dagger D
\ee
and $A_\mu,~\lambda,~D$ have R-charge 0,1,0, repectively.

Nelson and Seiberg discussed the relation between R symmetry and SUSY breaking \cite{Nelson}.
They showed under the condition 
\begin{itemize}
\item[i)]
Superpotential is generic, and
\item[ii)] low energy theory can be described by a supersymmetric Wess-Zumino model
\end{itemize}
that
\begin{itemize}
\item[a)]
R symmetry is necessary for SUSY breaking, and
\item[b)]
spontaneous R symmetry breaking is sufficient for SUSY breaking.
\end{itemize}

Thus if we have no U(1) symmetry we have appropriate SUSY vacuum, that is,
U(1) symmetry is necessary for SUSY breaking (condition (a)).

If there is U(1) symmetry and it is spontaneously broken, SUSY is automatically broken (condition (b)).

So the problem is how to impose $U(1)_R$ symmetry in superpotential of GUT.

Reflecting these situations, Ratz et al. \cite{Ratz} concluded that no MSSM model with either a ${\bf Z}_{M\geq 3}^R$ or U(1)$_{R}$ symmetry can be completed by a four dimensional GUT in the ultraviolet.
The essential point is explained for SU(5) GUT as follows.
$SU(5)\times Z_M^R$ is broken to the SM$\times  Z_M^R$ by the vev of the SM singlet of ${\bf 24}$.
${\bf 24}$ has zero R-charge since $Z_M^R$ is unbroken, and 
\be
{\bf 24}={\bf (8,1)}_0\oplus {\bf (1,3)}_0\oplus {\bf (1,1)}_0\oplus {\bf (3,2)}_{-5/6}\oplus {\bf (\overline{3},2)}_{5/6}.
\ee
Here $\la {\bf (1,1)}_0\ra\neq 0$, and ${\bf (3,2)}_{-5/6}$ and ${\bf (\overline{3},2)}_{5/6}$ get absorved to the longitudinal part of gauge bosons. The remaining ${\bf (8,1)}_0$ and ${\bf (1,3)}_0$ must be massive and therefore require mass term
$m{\bf 24}\times {\bf 24}$. However, it is prohibited because ${\bf 24}$ has 0 R-charge but superpotential must be 2 R-charge.
this is the case for more general mutiplet and more general gauge group including SO(10), The detail should be referred with \cite{Ratz}.
On the other hand in the case of Pati-Salam case,
PS group to the SM need to reduce rank by one, which is done by (4,1,2) and break B-L
quantum number and there give rise to no problem.
Therefore, the minimum group subject to no-go theorem is SU(5).

Of course there are a loophole of this no-go theorem.
For instance it is for meta-stable supersymmetry breaking vacuum, where $U(1)_R$ is broken explicitly \cite{Intrigator}.
That is, let us consider 
\be
W=-k\Phi_1+m\Phi_2\Phi_3+\frac{y}{2}\Phi_1\Phi_3^2
\ee.
which is $U(1)_R$ symmetric with R-charge, $R_{\Phi_1}=R_{\Phi_2}=2, R{\Phi_3}=0$.
\be
\Delta W=\frac{1}{2}\epsilon m\Phi_2^2,
\ee
where $\epsilon$ is a small dimensionless parameter. Thus we must explain this time why $\epsilon$ is so small to satisfy longevity of metastable state $\Phi_1=\Phi_2=\Phi_=0$ and we do not adopt this scenario.

On the otherhand, no-go theorem can not be applied in an extra dimensions, where new ways of GUT symmetry breaking mechanisms appear \cite{Witten} \cite{Breit} \cite{Kawamura}. This is one of very strong motivations to proceed to extra dimension.

We may consider \bref{U1R} from string theory.
In string theory \cite{Polchinski}, it has originally global space-time SO(10) symmetry and is broken to $SO(4)\times SO(6)$ in 4D.
This SO(6) is isomorphic to SU(4).
The spinor in ten space-time dimensions has $16_L+16_R$ components. (Do not confuse with flabour group so far discussed.)  
In the splitting from 10 to (4+6) dimensions, this spinor is divided into four 4-component spinor, $\theta_a^{(i)}, \theta_{\dot{a}}^{(i)}~~(a=1,2),~~i=1,2,3,4$
So there is $SU(4)_R$ transformation
\be
\theta'^{(i)}=U^i_j\theta^{(l)}.
\ee
\section{SO(10) GUT in 5D}
From this chapter we will realize the new model compatible with No-Go theorem discussed in the last part of 
previous chapter.
\subsection{Model Setup}
The model is described in 5D and 
 the fifth dimension is compactified 
 on the orbifold $S^1/{Z_2 \times Z_2^\prime}$. 
A circle $S^1$ with radius $R$ is divided by 
 a $Z_2$ orbifold transformation $y \to -y$ 
 ($y$ is the fifth dimensional coordinate $ 0 \leq y < 2 \pi R$)
 and this segment is further divided by a $Z_2^\prime$ transformation 
 $y^\prime \to -y^\prime $ with $y^\prime = y + \pi R/2$. 
There are two inequivalent orbifold fixed points at $y=0$ and $y=\pi R/2$. 
Under this orbifold compactification, a general bulk wave function 
 is classified with respect to its parities,  
 $P=\pm$ and $P^\prime=\pm$, under $Z_2$ and $Z_2^\prime$, respectively.

Assigning the parity ($P,P^\prime $) 
 the bulk SO(10) gauge multiplet suitably, 
 only the PS gauge multiplet has zero-mode 
 and the bulk 5D N=1 SUSY SO(10) gauge symmetry is broken 
 to 4D N=1 SUSY PS gauge symmetry \cite{Fuku8}. 
Since all vector multiplets has wave functions  
 on the brane at $y=0$, SO(10) gauge symmetry is respected there, 
 while only the PS symmetry is on the brane at $y=\pi R/2$ (PS brane). 

Its Yukawa coupling is given by
\bea
W_Y&=& Y_{1}^{ij} F_{Li} F_{Rj}^c H_1 
+\frac{Y_{15}^{ij}}{M_5} F_{Li} F_{Rj}^c 
 \left(H_1^{\prime} H_{15} \right) \nonumber\\ 
&+&\frac{Y_R^{ij}}{M_5} F_{Ri}^c F_{Rj}^c 
 \left(\phi \phi \right)  
 +\frac{Y_L^{ij}}{M_5} F_{Li}F_{Lj} 
 \left(\overline{H_L} \overline{H_L} \right), 
\label{Yukawa}
\eea 
Here the notations are as follows: $M_5$ is the 5D Planck scale. 
$F_{Li}$ and $F_{Ri}^c$ are matter multiplets 
 of i-th generation in $({\bf 4, 2, 1})$ and $({\bf \bar{4}, 1, 2})$ 
 representations, respectively. $H_1=({\bf 1},{\bf 2},{\bf 2})$,  
$H_1'=({\bf 1},{\bf 2},{\bf 2})'$, $H_{15}=({\bf 15},{\bf 1},{\bf 1})_H$,~
$H_6=({\bf 6},{\bf 1},{\bf 1})_H$, $\phi=({\bf 4},{\bf 1},{\bf 2})$, $\overline{\phi}=(\overline{{\bf 4}},{\bf 1},{\bf 2})$, $H_L=({\bf 4},{\bf 2},{\bf 1})_H$, $\overline{H_L}=(\overline{{\bf 4}},{\bf 2},{\bf 1})_H$ are Higgs multiplets.

The product, $H_1^{\prime} H_{15}$, effectively works 
 as $({\bf 15},{\bf 2},{\bf 2})_H$, 
 while $\phi \phi$ and $\overline{H_L}\overline{H_L}$ 
 effectively work as $({\bf 10},{\bf 1},{\bf 3})$ and 
 $(\overline{{\bf 10}},{\bf 3},{\bf 1})$, respectively, 
 and are responsible for the left- and the right-handed 
 Majorana neutrino masses. 
Providing VEVs for appropriate Higgs multiplets, 
 fermion mass matrices are obtained.

\begin{eqnarray}
 M_u &=& c_{10} M_{1,2,2}+ c_{15} M_{15,2,2},~~
 M_d = M_{1,2,2} + M_{15,2,2} \; ,   
 \nonumber \\
 M_D &=& c_{10} M_{1,2,2} - 3 c_{15} M_{15,2,2},~~
 M_e = M_{1,2,2} - 3 M_{15.2,2} \; , 
  \\
 M_L &=& c_L M_{10,3,1},~~ 
 M_R = c_R M_{10,1,3} \; . \nonumber
\label{massmatrix2}
\end{eqnarray}  

Two important remarks are in order. \\
1. $M_{15,2,2}$ is, in general, not symmetric unlike $M_{126}$. However, we imposed the
L-R symmetry ${\bf 4,1,2}\leftrightarrow {\bf \bar{4},2,1}$, which implies that bothy  $M_{1,2,2}$ and $M_{15,2,2}$ matrices are symmetric and mass structure of charged Fermions and Dirac neutrino is same as that in SO(10).\\
2. $M_L$ and $M_R$ are independent on those of the charged Fermions and the Dirac neutrino unlike the SO(10) case (See Eq.\bref{massmatrix}).
So the precise data fitting becomes possible without changing $Y_\nu$.
This is very important especially for LFV and leptogenesis.

$H_6$ is necessary to make the color triplet heavy. However, there arises no Doublet-Triplet problem since they are not involved in the same multiplet. There are sufficient numbers of free parameters 
 to fit all the observed fermion masses and mixing angles. 

\subsection{SUSY breaking and Dark Matter}
In the orbifold GUT model, we assume that 
 the GUT model takes place at some high energy 
 beyond the compactification scale.  
For the theoretical consistency of the model, 
 the gauge coupling unification should be realized 
 at some scale after taking into account 
 the contributions of Kaluza-Klein modes 
 to the gauge coupling running.

In our setup, the evolution of gauge coupling 
 has three stages, $G_{321}$ (SM+MSSM), $G_{422}$ (whose energy scale is $v_{PS}$) 
 and $M_c =1/R$.
From the model setting we adopted gaugino mediation mechanism as SYSY breaking scenario.
First we simply assumed $v_{PS}=M_c$ \cite{Fuku8}.
In this case, stau becomes the lightest SUSY particle (LSP).

In order to remedy this trouble we next considered $M_c>v_{PS}$ and showed that neutralino becomes the LSP at \cite{Fuku10}
\bea 
 M_c = 2.47 \times v_{PS} = 2.95 \times 10^{16} \;  \mbox{GeV}. 
\eea 

We gives the gauge coupling running in both cases.
\begin{figure}[ht]
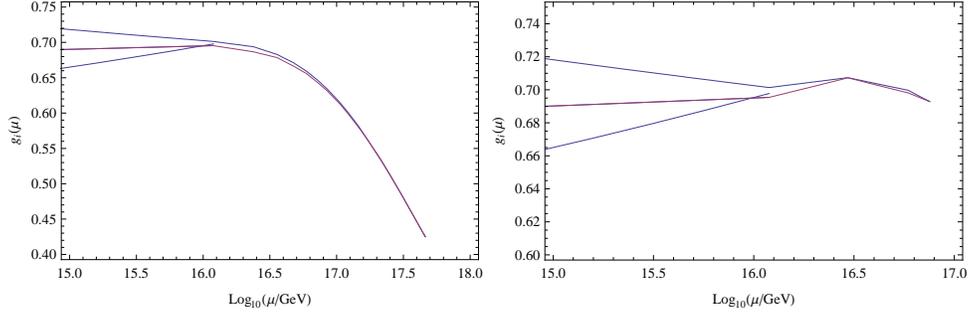

\includegraphics[scale=0.6]{Fig1.eps}
\includegraphics[scale=0.6]{Fig1stau.eps}
\caption{Left pannel:
Gauge coupling unification in the left-right symmetric case, 
 taken from \cite{Fuku8}. 
Each line from top to bottom corresponds to 
 $g_3$, $g_2$ and $g_1$ for $ \mu < M_c=v_{\rm PS}$, 
 while  $g_3=g_4$ and $g_2=g_{2R}$ for $ \mu > M_c=v_{\rm PS}$. 
 Right pannel:Gauge coupling unification for $M_c>v_{PS}$ from \cite{Fuku10}. 
Each line from top to bottom corresponds to 
 $g_3$, $g_2$ and $g_1$ for $ \mu < v_{PS}$,
 while  $g_3=g_4$ and $g_2=g_{2R}$ for $ \mu > v_{PS}$. 
Here, we have taken $M_c= 2.47 \times v_{PS}$.
}
\label{GCU2}
\end{figure}

\subsection{Confrontation with Cosmology--Smooth hybrid inflation}
Original single-field inflaton theory suffered from fine tuning problem though observational check is due to its prediction on non-Gaussianity $f_{NL}\approx 0.02$ \cite{Komatsu}.
In this susection we discuss the smooth hybrid inflation \cite{Lazarides} in the context of a simple supersymmetric SO(10) GUT in 5D orbifold \cite{Fuku9}. (For another hybrid model to solve monopole problem (shifted hybrid inflation), see \cite{Shafi}.)
Let us consider the superpotential for the smooth hybrid inflation
\bea
 W= S \left( -\mu^2+\frac{(\bar{\phi} \phi)^2}{M^2} \right).
\label{smooth}  
\eea
Here $\phi$ and $\overline{\phi}$ are defined in \bref{Yukawa} and we have omitted possible ${\cal O}(1)$ coefficients. 
SUSY vacuum conditions lead to non-zero VEVs for 
 $\langle \phi \rangle = \langle \bar{\phi} \rangle = \sqrt{\mu M}$, 
 by which the PS symmetry is broken down to the SM one, and thus  
\bea 
  v_{\rm PS} = \sqrt{\mu M}.   
\eea 
We evaluated the spectral index, 
 the tensor-to-scalar ratio and the running of the spectral index: 
\bea 
   0.963 \leq  & n_{\rm s} & \leq 0.968,   \nonumber \\ 
  4.0 \times 10^{-7} \geq  & r &   \geq  3.1 \times 10^{-7} ,
  \nonumber \\ 
  -8.4 \times 10^{-4} \leq & \alpha_{\rm s} & \leq -6.1 \times 10^{-4} 
\eea  
 for 1 MeV $\leq T_{\rm rh} \leq 10^7$ GeV. 
The tensor-to-scalar ratio and the running of the spectral index 
 are negligibly small. 
These results are consistent with the WMAP 5-year data \cite{WMAP}: 
 $n_{\rm s} = 0.960^{+0.014}_{-0.013}$, 
 $ r <  0.2$ (95\% CL) 
 and $ \alpha_{\rm s} = -0.032^{+0.021}_{-0.020}$ (68\% CL) 
 (consistent with zero in 95\% CL). 
We also discussed on the non-thermal leptogenesis \cite{Fuku11}.
As the mass relation between charged fermions are same as minimal SO(10) and we can use
$Y_\nu$ of Eq.\bref{Ynu}. whereas we can not reproduce MNS uniquely grom model and assumed tri-bimaximal model \cite{Harrison}. 
The resultant baryon asymmetry is obtained as a function of the lightest mass eigenvalue of the light neutrinos, and we find that a suitable amount of baryon asymmetry of the universe can be produced in the normal hierarchical case, while in the inverted hierarchical case the baryon asymmetry is too small to be consistent with the observation.

Thus the advantageous points of minimal SO(10) are succeeded to the SO(10) model in 5D, which goes over the mismatches with observations as well as the conceptual trouble indicated by several no-go theorems \cite{Opening}.

\begin{theacknowledgments}
  The works discussed in this talk are mainly based on the collaboration with N.Okada, S.Meljanac, A.Ilakovac, T.Kikuchi. This work is partly supported by the Grant-in-Aid for Scientific Research from the Ministry of Education, Science and Culture of Japan (No.020540282 and No. 21104004). 
\end{theacknowledgments}

\end{document}